\documentclass{article}
\usepackage{amssymb}
\usepackage{float}
\usepackage[letterpaper,top=2cm,bottom=2cm,left=3cm,right=3cm,marginparwidth=1.75cm]{geometry}

\usepackage{amsmath}
\usepackage{graphicx}
\usepackage{CJKutf8}
\usepackage[colorlinks=true, allcolors=blue]{hyperref}

\title{User Behavior Analysis and Clustering in a MMO Mobile Game: Insights and Recommendations}

\author{
    Yang Qiu \\
    \texttt{yaq001@ucsd.edu} \\
    University of California, San Diego
    \and
     Yuxin Gong \\
     \texttt{gongyx@seu.edu.cn} \\
    Southeast University
    \and
    Guanliang Liu \\
    Southeast University
}

\begin{document}
\begin{CJK}{UTF8}{gbsn}
\maketitle

\begin{abstract}
This study presents a comprehensive analysis of user behavior and clustering in a MMO mobile game, a popular mobile battle royale game, employing temporal and static data mining techniques to uncover distinct player segments. Our methodology encompasses time series K-means clustering, graph-based algorithms (DeepWalk and LINE), and static attribute clustering, visualized through innovative hybrid charts. Key findings reveal significant variations in player engagement, skill levels, and social interactions across five primary user segments, ranging from highly active and skilled players to inactive or new users. We also analyze the impact of external factors on user retention and the network structure within clusters, uncovering correlations between cluster cohesion and player activity levels. This research provides valuable insights for game developers and marketers, offering data-driven recommendations for personalized game experiences, targeted marketing strategies, and improved player retention in online gaming environments.

\end{abstract}

\section{Data Overview \& Overall Analysis}
\subsection{Statistical Analysis}

\subsection{Key Opinion Leader Mining}
In order to explore whether there are common characteristics of Key Opinion leaders (Kols) in players, this study uses PG value calculated by PageRank algorithm as the influence measurement standard of user nodes.
\begin{figure}[H]
    \centering
    \includegraphics[width=0.8\textwidth]{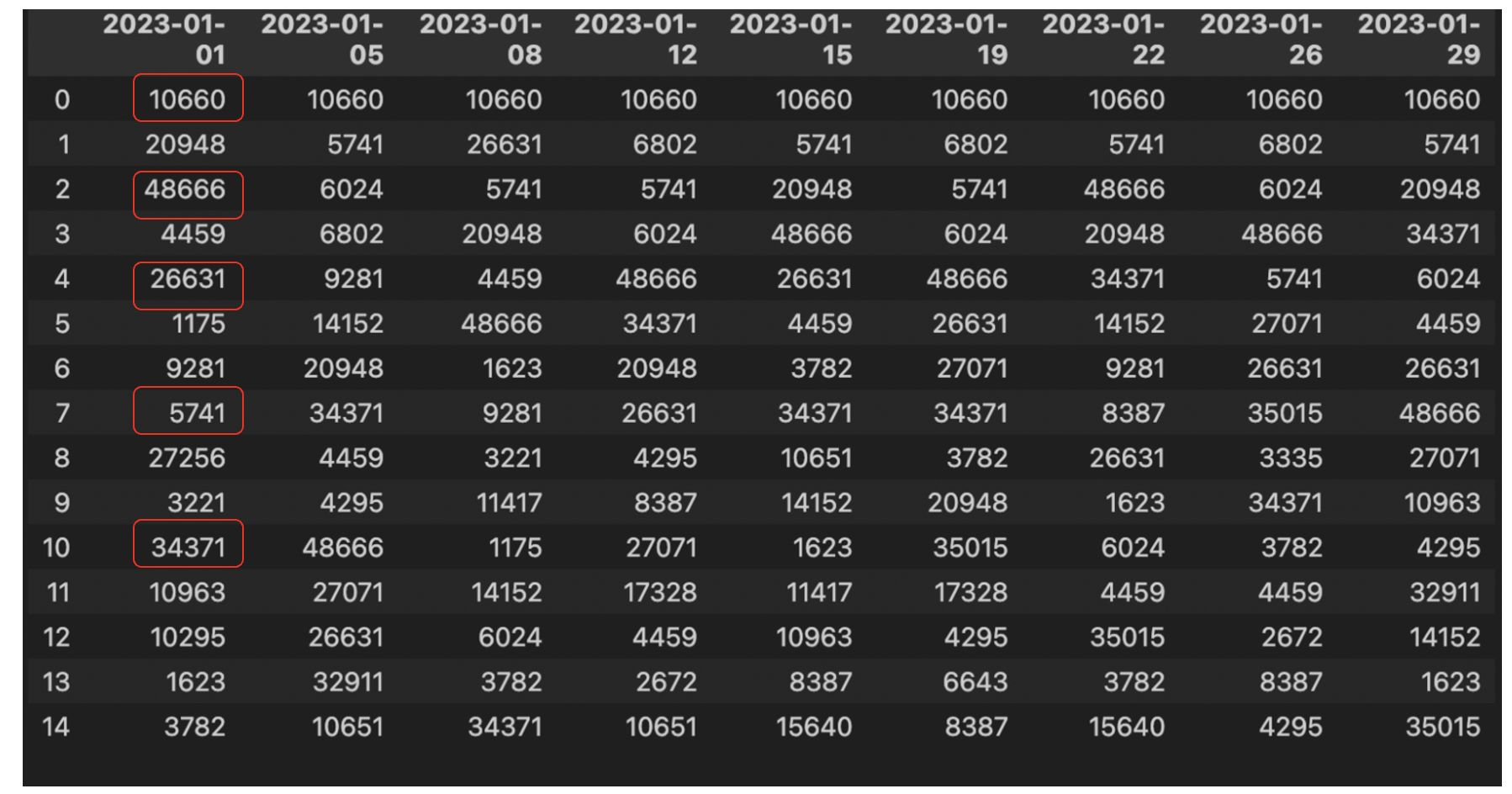}
    \caption{ID(Masked ) of Kols at different time points.}
    \label{fig:kol}
\end{figure}

As shown in Figure \ref{fig:kol}, there exist 5 users in the Top15-KOL user lists all the time. By analyzing the personal static attributes of these users, it is easy to find that they have the following commonalities：
\begin{itemize}

\item  All belonging to the game resident players (non-comeback/non-new registered users)
\item  Logining frequently, friends are almost full list
\item  Higher segment, higher level
\item  Higher average willingness to pay
\end{itemize}

\section{Data Clustering (Temporal/Dynamic Data Mining)}

In the study of user behavior in gaming environments, particularly in a MMO mobile game, clustering users based on their interaction patterns can reveal distinct player types and preferences. This analytical approach facilitates the development of personalized game experiences and effective marketing strategies.

Features such as acquisition (\texttt{diamond\_add\_1week}), and other relevant metrics are captured. Data from different files are merged to form a comprehensive dataset, where each user's data spans multiple time points.

The features are processed to create a multivariate time series for each user. This involves pivoting the dataset to format the data such that each row corresponds to a user and each column to a specific feature at a given time point, filling any missing values with zeros to maintain consistency.

\subsection{Feature Engineering}

Additional features, such as the \texttt{mode\_choice\_ratio}, which represents the preference of users between different game modes, are computed from the raw data to provide deeper insights into user behavior. This ratio is calculated by dividing the number of times a user engages in a funny-mode game mode by the total number of games played, filled with zero in cases of undefined expressions (such as division by zero).

The selection of features for clustering was guided by a comprehensive analysis aimed at identifying attributes with significant predictive power and minimal multicollinearity. The attributes selected include team leadership frequency (\texttt{carteam\_leader\_num}), win rate (\texttt{chicken\_rate}), weekly diamond acquisition (\texttt{diamond\_add\_1week}), game mode choice ratio (\texttt{mode\_choice\_ratio}), return player indicator (\texttt{is\_comeback}), average damage (\texttt{avg\_damage}), recruitment numbers (\texttt{recruit\_num}), registration status (\texttt{is\_register}), number of platform friends (\texttt{friend\_num\_plat}), and average health recovery times (\texttt{avg\_healtimes}). These features were chosen based on their relevance to player behavior and game engagement.

To refine the selection of attributes for clustering, a composite score was calculated for each attribute based on its normalized Average Correlation, Variance Inflation Factor (VIF), and contributions from Principal Component Analysis (PCA). The score is designed to balance the relevance, uniqueness, and overall contribution of each attribute to the dataset's variance, as explained below:

\begin{itemize}
    \item \textbf{Normalized Average Correlation} quantifies the direct relationship each attribute has with the target variable, scaled between 0 and 1 for comparability. Higher values indicate stronger relationships. The normalization is calculated as:
    \[
    \text{Normalized Correlation} = \frac{\text{Average Correlation} - \min(\text{Average Correlation})}{\max(\text{Average Correlation}) - \min(\text{Average Correlation})}
    \]
    
    \item \textbf{Normalized VIF} assesses how much the attribute's variance is inflated due to linear dependencies with other predictors. Lower VIF values are preferable as they indicate lower multicollinearity. The normalization is given by:
    \[
    \text{Normalized VIF} = \frac{\text{VIF} - \min(\text{VIF})}{\max(\text{VIF}) - \min(\text{VIF})}
    \]
    
    \item \textbf{Normalized PCA Contributions} measure the extent to which each attribute influences the principal components of the dataset. This metric helps in understanding the attribute's importance in capturing the variance within the dataset. The normalization is performed similarly:
    \[
    \text{Normalized PCA Contributions} = \frac{\text{PCA Contributions} - \min(\text{PCA Contributions})}{\max(\text{PCA Contributions}) - \min(\text{PCA Contributions})}
    \]

\end{itemize}

The composite score for each attribute is then computed as:
\[
\text{Score} = (\text{Normalized Correlation}) + (1 - \text{Normalized VIF}) + (1 - \text{Normalized PCA Contributions})
\]
This formula integrates the three normalized metrics by prioritizing attributes with low multicollinearity, high predictive power, and significant contributions to the principal components. Attributes with lower scores are typically considered less suitable due to either high redundancy, low relevance, or minimal contribution to the dataset's variance.

The following table lists the attributes sorted by their composite scores, highlighting those that provide the most balanced insight into player behaviors without redundancy:

\begin{table}[h]
\centering
\caption{Top10 Attributes and their corresponding statistics}
\label{tab:attributes_simplified}
\begin{tabular}{lcccc}
\hline
\textbf{Attribute} & \textbf{Norm. PCA Contrib.} & \textbf{Norm. VIF} & \textbf{Norm. Corr.} & \textbf{Score} \\

\hline
carteam\_leader\_num & 0.396105 & 0.000000 & 0.000000 & 2.603895 \\
chicken\_rate & 0.271855 & 0.061501 & 0.330336 & 2.336308 \\
diamond\_add\_1week & 0.551655 & 0.001751 & 0.126842 & 2.319752 \\
mode\_choice\_ratio & 0.684337 & 0.021597 & 0.125096 & 2.168969 \\
is\_comeback & 0.637972 & 0.004038 & 0.189655 & 2.168335 \\
avg\_damage & 0.000000 & 0.148985 & 0.710221 & 2.140794 \\
recruit\_num & 0.744416 & 0.006931 & 0.152095 & 2.096558 \\
is\_register & 0.926395 & 0.002707 & 0.048149 & 2.022748 \\
friend\_num\_plat & 0.904019 & 0.005362 & 0.074169 & 2.016449 \\
avg\_healtimes & 0.525981 & 0.014838 & 0.479710 & 1.979472 \\
\hline
\end{tabular}
\end{table}
Excluded attributes, such as map-specific usage and round counts, are omitted from this table due to their identified lower impact on the overall analysis based on their scores.

\subsection{Clustering with TimeSeriesKMeans}
Time Series K-Means clustering is a specialized version of the K-Means algorithm designed to handle time-series data. This algorithm clusters data considering the temporal correlation between observations which is crucial for time-series analysis. Each data point in a time-series dataset is a vector of observations across multiple time points.

Given a set of $N$ time-series data points, each represented as $X_i = \{x_{i1}, x_{i2}, \ldots, x_{iT}\}$ where $x_{it}$ is a vector in $\mathbb{R}^d$ representing the data at time $t$ with $d$ features, the goal is to partition the $N$ time-series into $K$ clusters. Each cluster is characterized by a centroid time-series $C_k = \{c_{k1}, c_{k2}, \ldots, c_{kT}\}$, also in $\mathbb{R}^d$.

The TimeSeriesKMeans algorithm aims to minimize the total within-cluster variance, defined as:
\[
S = \sum_{k=1}^{K} \sum_{X_i \in C_k} \sum_{t=1}^{T} \| x_{it} - c_{kt} \|^2
\]

Where:
\begin{itemize}
    \item $K$ is the number of clusters.
    \item $C_k$ is the set of time-series in cluster $k$.
    \item $x_{it}$ is the feature vector of the $i$-th time-series at time $t$.
    \item $c_{kt}$ is the centroid vector of the $k$-th cluster at time $t$.
    \item $\|\cdot\|$ denotes the Euclidean distance.
\end{itemize}

The algorithm iterates through the following steps:
\begin{enumerate}
    \item \textbf{Initialization}: Select $K$ initial centroids either randomly or based on a heuristic.
    \item \textbf{Assignment step}: Assign each time-series $X_i$ to the cluster $C_k$ whose centroid $C_k$ minimizes the distance $\sum_{t=1}^{T} \| x_{it} - c_{kt} \|^2$.
    \item \textbf{Update step}: Update each cluster centroid $C_k$ to be the mean of all time-series assigned to it, calculated separately for each time point $t$.
    \item \textbf{Convergence check}: Repeat the assignment and update steps until the centroids do not change significantly, indicating convergence.
\end{enumerate}

This clustering method is particularly useful for identifying patterns in temporal data where the sequence and timing of the observations are crucial. It has been effectively applied in various fields such as economics, meteorology, and behavioral analytics.

\subsection{Visualization of Clustering with PCA and t-SNE}

Principal Component Analysis (PCA) and t-Distributed Stochastic Neighbor Embedding (t-SNE) are two widely used techniques for dimensionality reduction and visualization of high-dimensional data. Both methods serve to simplify the complexities of multivariate data, making it easier to interpret clustering results visually.

\subsubsection{PCA - Clustered Data Visualization}

PCA reduces the data by projecting it onto a set of orthogonal axes that capture the directions of maximum variance. This transformation results in a new coordinate system where the first principal component captures the greatest variance, followed by the second, and so on. By reducing the data to the first two principal components, we can visualize the clusters in a two-dimensional space.

\begin{figure}[H]
    \centering
    \includegraphics[width=0.8\textwidth]{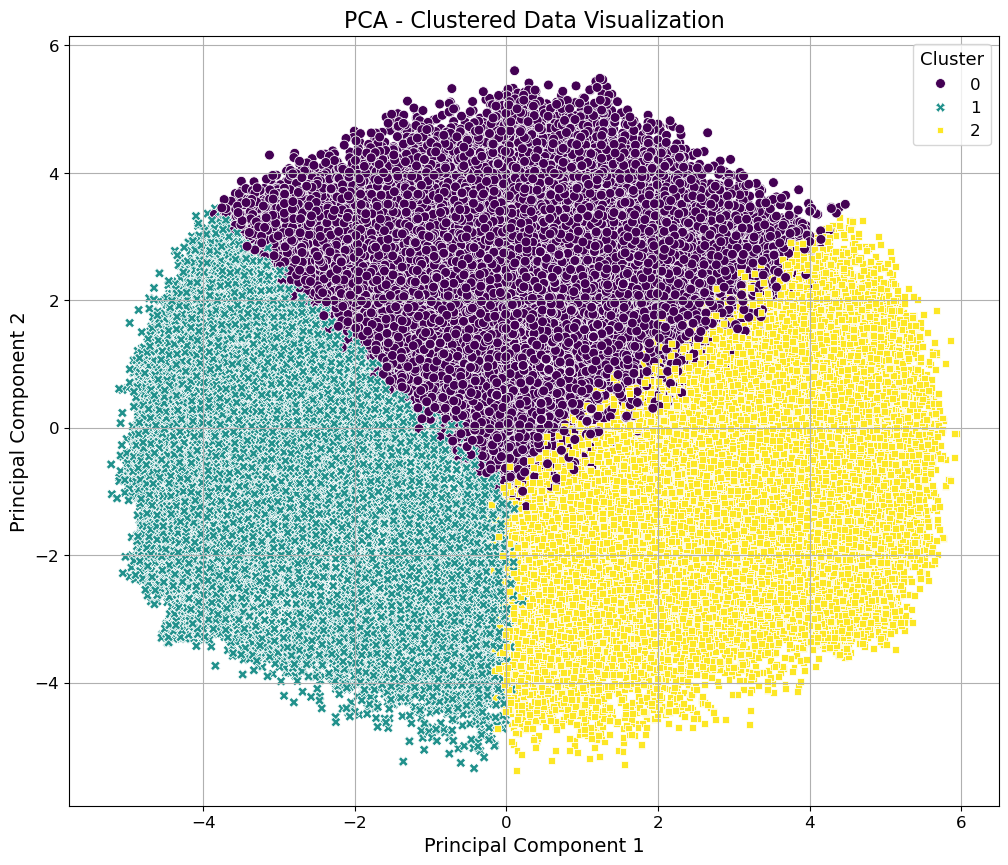}
    \caption{PCA visualization of clustered data.}
    \label{fig:pca}
\end{figure}

In Figure \ref{fig:pca}, each point represents a user, colored according to their cluster assignment. 

\subsubsection{t-SNE - Clustered Data Visualization}

Unlike PCA, t-SNE is a nonlinear technique that focuses on preserving the local structure of the data, making it particularly effective at separating clusters in a high-dimensional space into a two-dimensional plot. t-SNE maps the high-dimensional data to a lower-dimensional space by optimizing the similarity between points, ensuring that points that are close in the high-dimensional space remain close in the low-dimensional representation.

\begin{figure}[H]
    \centering
    \includegraphics[width=0.8\textwidth]{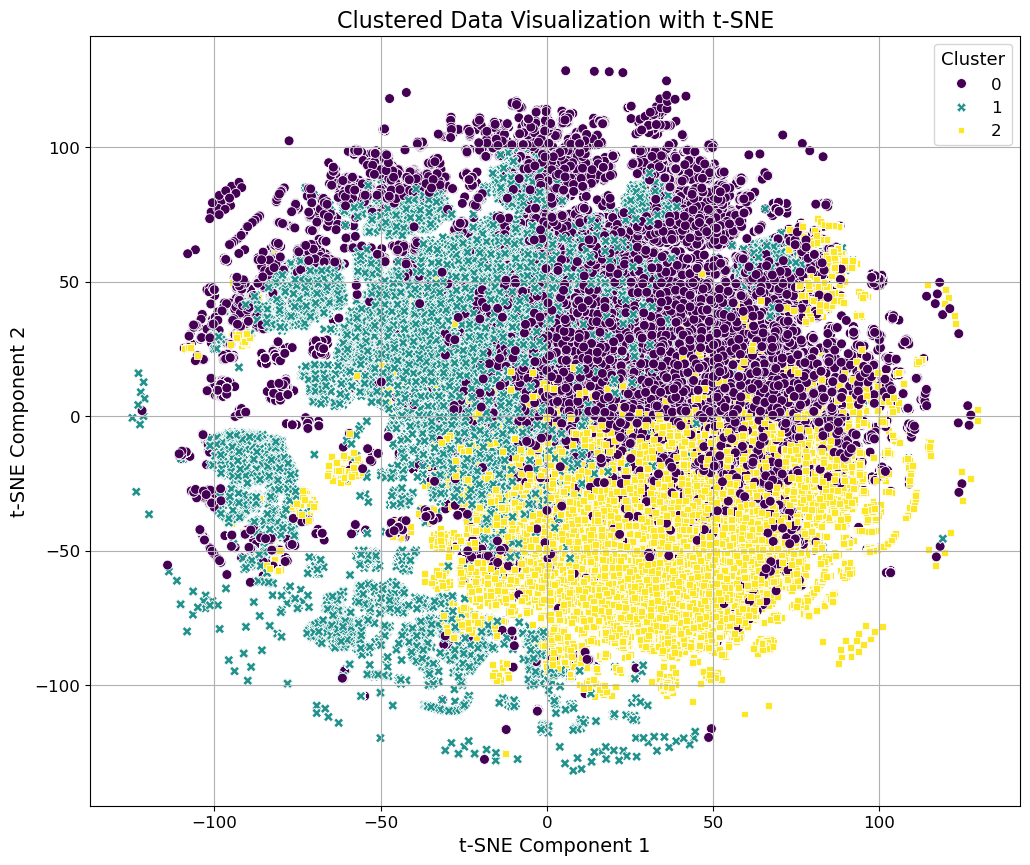}
    \caption{t-SNE visualization of clustered data.}
    \label{fig:tsne}
\end{figure}

As shown in Figure \ref{fig:tsne}, t-SNE provides a more intuitive visualization of the data clusters. The visualization illustrates strong homogeneity and good separation across clusters, which highlights the effectiveness of both the clustering algorithm and t-SNE in uncovering inherent patterns in the dataset.

\section{Analysis of Cluster 1's User Profile and Impact of Upcoming Game Content}

\begin{figure}[H]
    \centering
    \includegraphics[width=\textwidth]{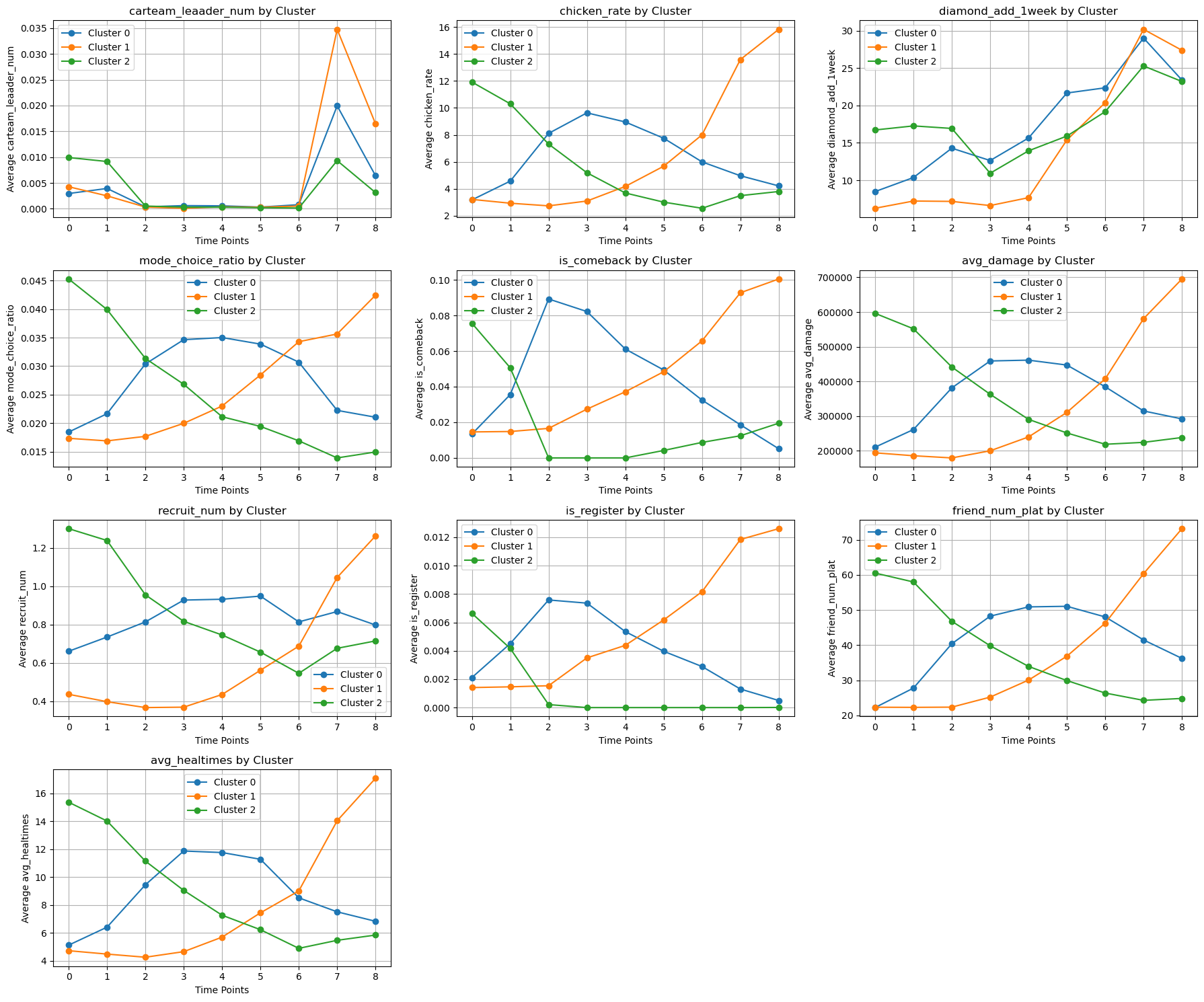}
    \caption{Trend Changes of Three Clusters.}
    \label{fig:tendency}
\end{figure}

Cluster 1's user profile initially exhibits the lowest metrics across all categories at the first time point, presenting a unique case within the data. Despite starting from the lowest baseline, this cluster demonstrates a remarkable trajectory of rapid improvement, surpassing other clusters and achieving the highest metrics by the final time point. This dramatic rise from last to first highlights the players' high levels of activity, competitiveness, and a strong ability to adapt and enhance their gameplay significantly over time. Players in this cluster are not only frequent participants but also continually advance their skills, in-game resources, and engage robustly in social interactions. They maintain numerous in-game friendships, often assume leadership roles, and show a keen interest in exploring new content and game mechanics.

For the second and third time points, there was a noticeable decline in new registrations and returning users. This section explores potential causes and implications of this trend, focusing on the mid-October server issues.

\begin{figure}[H]
    \centering
    \includegraphics[width=0.9\textwidth]{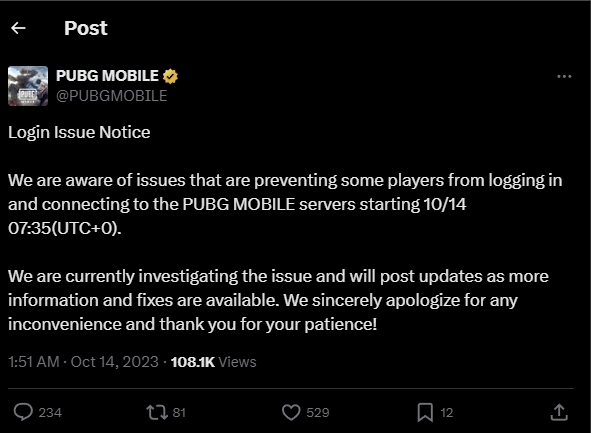}
    \caption{PUBG Official Statement on Login Issues.}
    \label{fig:Login}
\end{figure}

\subsection{Impact of Server Issues}
On October 14, 2023, PUBG Mobile experienced significant server issues, impacting player connectivity and potentially deterring new and returning users. Although the issues were resolved within approximately 4 hours, widespread disruptions had already been reported by users at least an hour before the company's public acknowledgment. This delay in response from the public relations team led to an accumulation of a substantial number of negative comments on social media platform X under posts related to the outage. The visible frustration and negative sentiment from this disruption, widely shared and discussed on these platforms, likely led to a temporary decrease in engagement. Specifically, the visibility of the problem, highlighted through a tweet by PUBG Mobile's official account with 108.1K views, could have influenced the community's perception, further discouraging participation during this period.

\subsection{User Confidence and Trust}
Trust in the game's reliability is critical for engaging new and returning players. Events like server downtimes, especially if occurring during a player's initial engagement, can erode this trust and discourage further investment in the game. Such incidents emphasize the importance of robust game infrastructure and responsive customer service to maintain player confidence and loyalty. Notably, even after the game's official announcement that the login issues had been resolved, comments on social media indicated that some players were still unable to log in. This discrepancy between the official statements and the players' actual experiences could lead to perceived mistrust and skepticism regarding the game's reliability and the transparency of the communication from the game's operators.

\subsection{Recommendations for Operational and Public Relations Teams}
Given the significant impact of server downtimes on player retention and acquisition, it is crucial for operational teams to implement more robust monitoring tools and quicker troubleshooting procedures to minimize downtime duration. Similarly, public relations teams should be prepared with rapid response strategies, including transparent and frequent communications with the player base during outages. Developing a comprehensive crisis management plan that includes proactive community engagement and clear, reassuring messaging can mitigate negative sentiment. Additionally, offering compensation or incentives to affected players can help restore trust and encourage continued engagement after the incident.

Future contingency planning should involve:
\begin{itemize}
    \item Establishing a dedicated rapid response team for immediate technical issues and communication.
    \item Regularly updating players on social media and other communication channels during the resolution process.
    \item Preparing fallback servers to reduce the load and switch traffic in case of failure.
    \item Conducting regular, scheduled stress tests on the infrastructure to ensure stability under high demand.
\end{itemize}

\section{Static Data Mining }
In this section, we will do some interesting analysis with the user data at time point 2023.10.22 as an example.
\subsection{Data Clustering on Graph Structure}
Considering that the dataset contains the social relationship graph of users, we can mine the structural features of the graph as the basis of user clustering.To achieve this, two graph mining algorithms will be referred to here.

\subsubsection{DeepWalk}
DeepWalk is a graph embedding algorithm based on random walks, introduced by Bryan Perozzi, Rami Al-Rfou, and Steven Skiena in 2014. Its primary objective is to map nodes from complex networks into a low-dimensional vector space, where these vectors capture the structural properties and semantic information between nodes. This representation proves valuable for network analysis tasks such as node classification, link prediction, and community detection. 

DeepWalk revolves around the concept of generating sequences (or "walks") through random walks on the graph, which are then used as inputs to word embedding techniques commonly employed in natural language processing, like Word2Vec, to learn node embeddings. Given that the social graph between users in this dataset is undirected, it is natural to contemplate employing the DeepWalk algorithm to generate user representations.

\begin{figure}[htbp]
  \centering
  \begin{minipage}[b]{0.45\textwidth}
    \includegraphics[width=\textwidth]{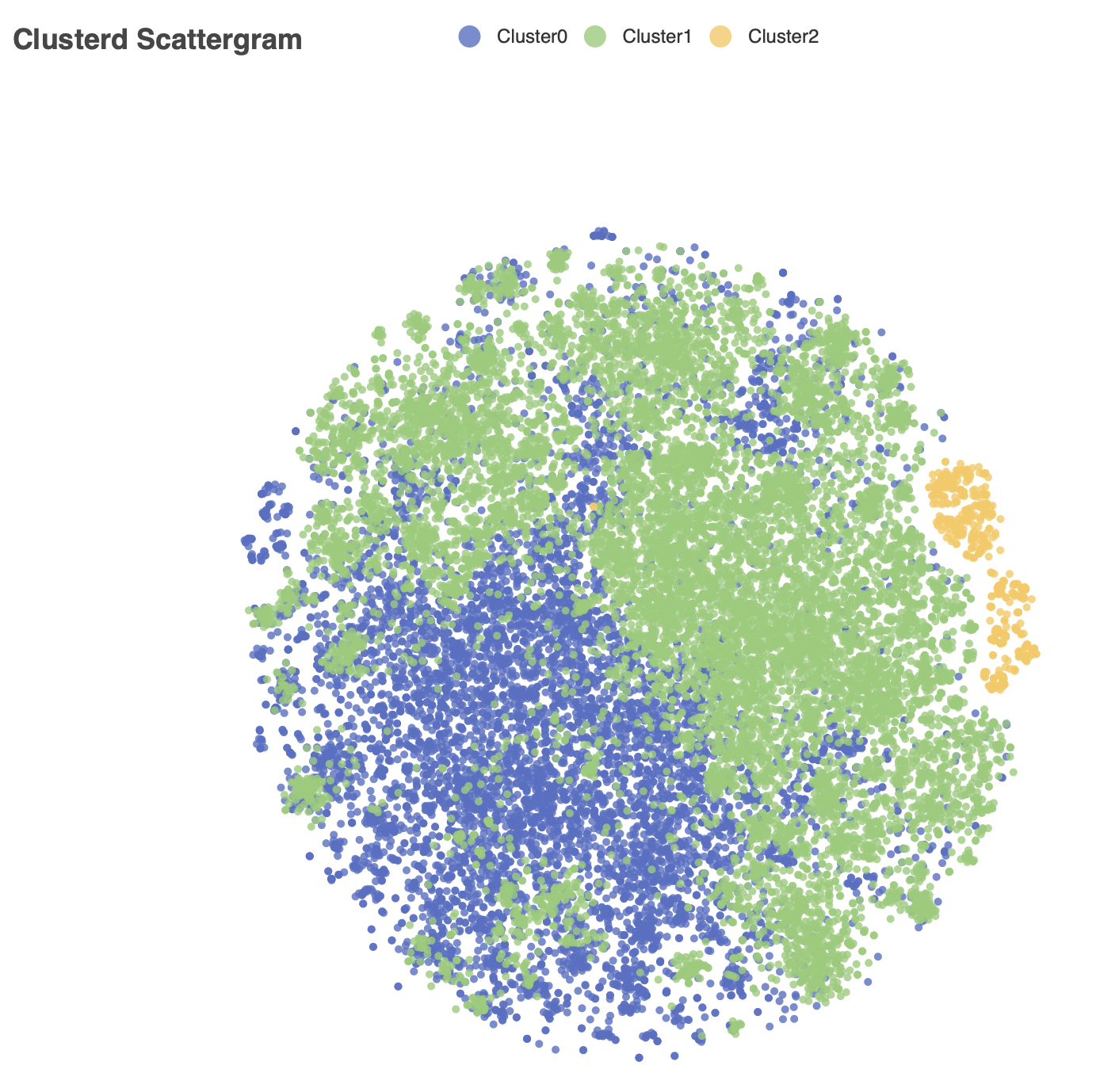}
    \caption{User clustering visualization with DeepWalk.}
    \label{fig:deep}
  \end{minipage}
  \hfill
  \begin{minipage}[b]{0.45\textwidth}
    \includegraphics[width=\textwidth]{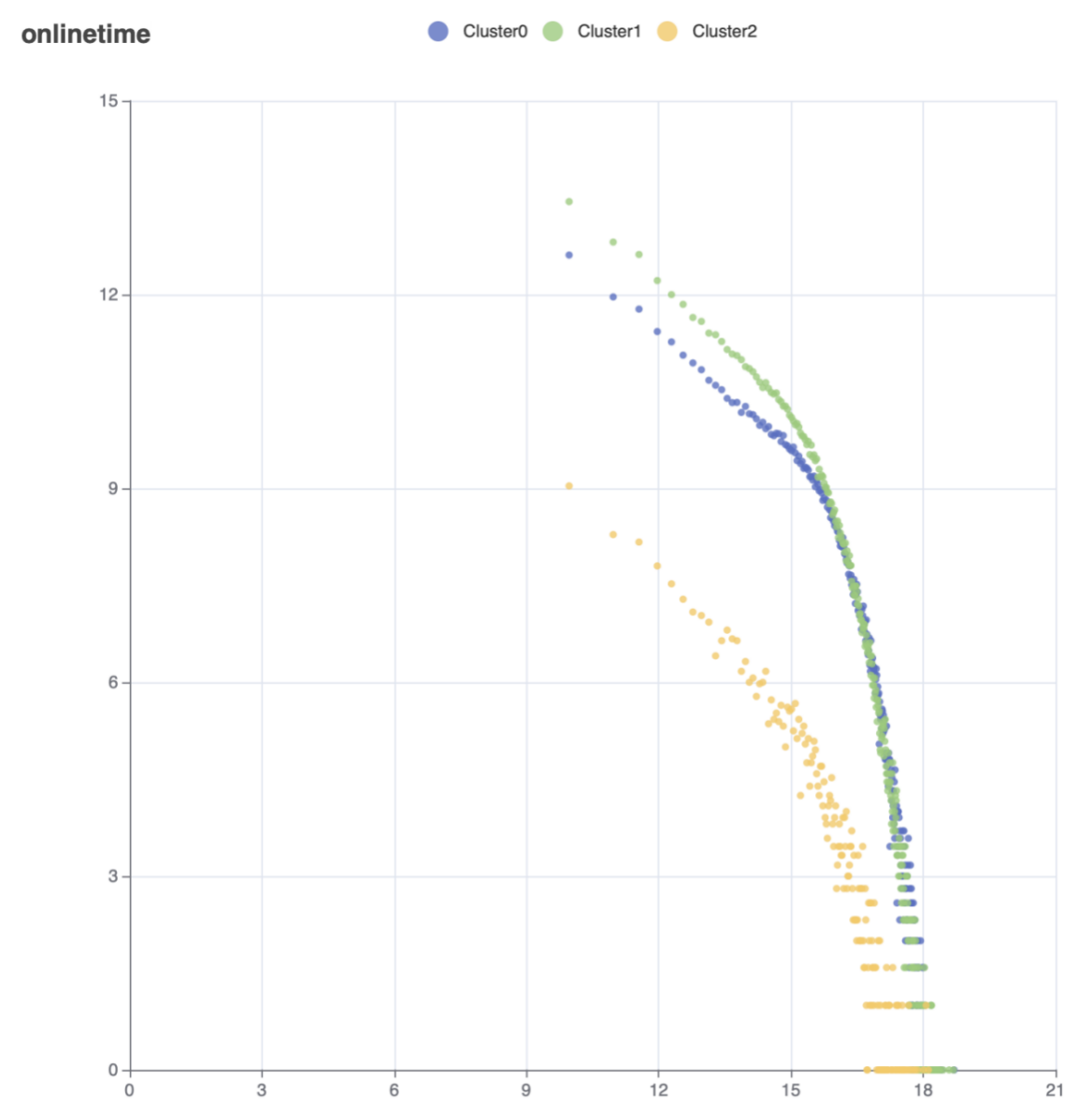}
    \caption{Online duration feature visualization with DeepWalk.}
    \label{fig:deep-on}
  \end{minipage}
\end{figure}

Figure \ref{fig:deep} is a clusterd scattergram after dimensionality reduction by t-SNE method. Figure \ref{fig:deep-on} shows the online duration distribution of different user groups after K-means clustering. Where, the X-axis represents the grouping of online duration. In this study, we take 1000 as the dividing unit. As the group number increases, the online duration increases correspondingly, and the Y-axis represents the total number of users who reach the corresponding online duration.

As shown in Figure \ref{fig:deep} and \ref{fig:deep-on}, the online duration distribution of different user clusters is distinct, which proves that using Deepwalk algorithm representation can filter inactive user groups from the graph structure.

\subsubsection{LINE}
LINE (Large-scale Information Network Embedding) is an algorithm designed for embedding structured data from networks such as social networks, semantic networks, or knowledge graphs. It aims to learn low-dimensional representations (embeddings) of nodes in these networks that capture both the complex network structures and underlying relationships. Unlike methods like DeepWalk, which rely on random walks, LINE specifically focuses on preserving two types of structural information in networks: First-order Proximity and Second-order Proximity. 

LINE achieves this by defining a joint probability model and optimizing node embeddings through a target function, typically negative log-likelihood. This approach enables LINE to efficiently learn high-quality node representations on large-scale networks, which are applicable to various downstream tasks including link prediction, node classification, and community detection. Here, we also try to use LINE algorithm to mine graph structure information to generate user characteristics.

\begin{figure}[H]
    \centering
    \includegraphics[width=0.6\textwidth]{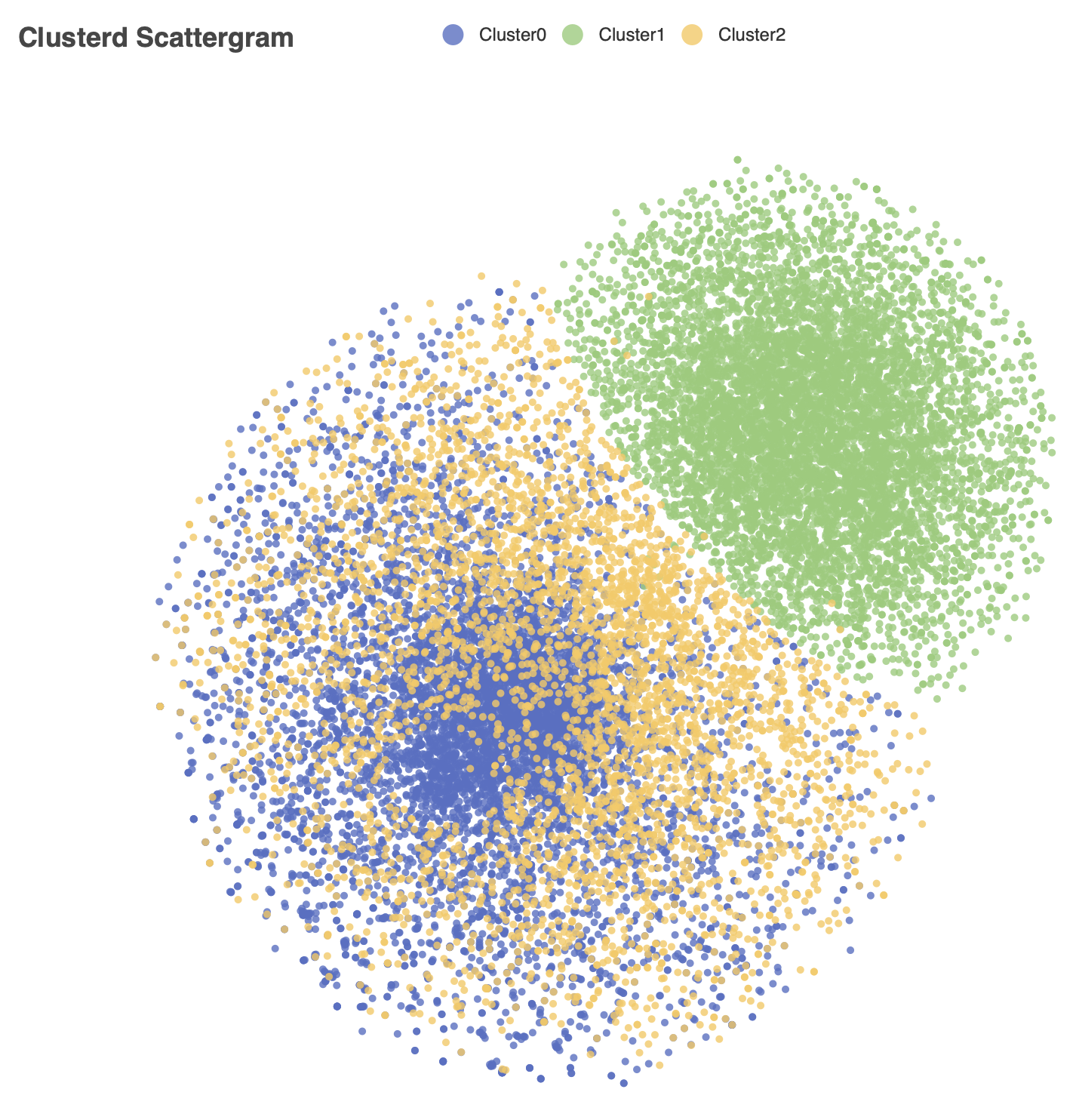}
    \caption{User clustering visualization with LINE.}
    \label{fig:line}
\end{figure}

\begin{table}[h]
\centering
\caption{User distribution}
\label{tab:line}
\begin{tabular}{lcc}
\hline
\textbf{Cluster} & \textbf{Proportion of Newly Registered Users} & \textbf{Proportion of Comeback Users} \\

\hline
0 & 0.2\% & 4.1\% \\
1 & \textbf{15.2\% }& 10.3\% \\
2 & 0.2\% & 4.4\% \\
\hline
\end{tabular}
\end{table}

Unlike Figure \ref{fig:deep}, Figure \ref{fig:line} can clearly filter out  individual user groups from the graph structure information, such as Cluster 1. By analyzing Table \ref{tab:line}  and Figure \ref{fig:line}, we can find that the proportion of new registered users in Cluster 1 far exceeds that of the other two user groups, which means that using the LINE algorithm representation may help us filter new registered user groups more easily.

\subsection{Data Clustering on User Static Attributes}
\begin{figure}[H]
    \centering
    \includegraphics[width=0.8\textwidth]{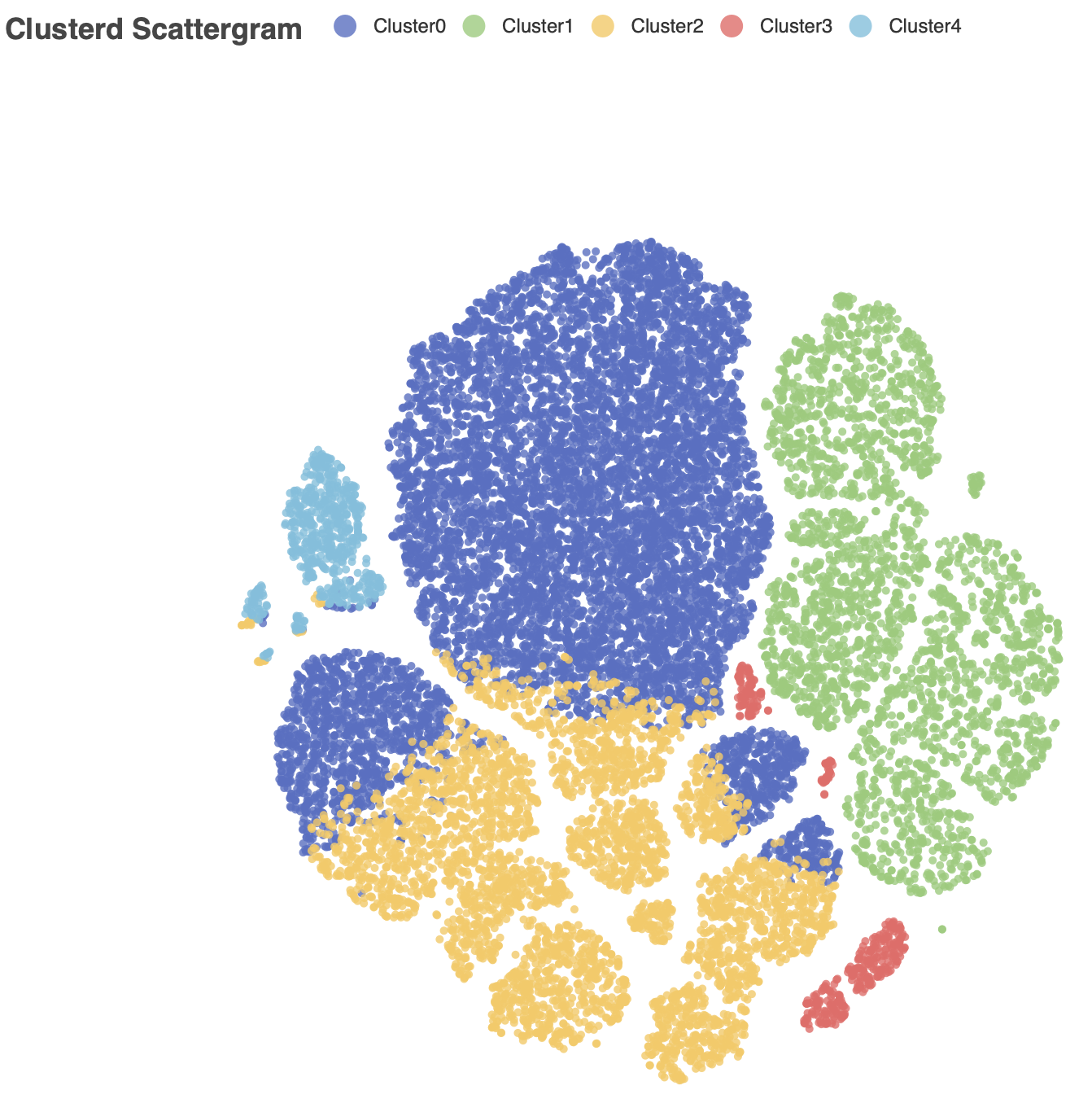}
    \caption{User clustering visualization with user static attributes.}
    \label{fig:static}
\end{figure}

In this section, static attributes with high orthogonality are selected for user clustering, and 5 user groups are formed are formed by K-means clustering. Similarly, we used T-SNE to visualize the distribution of different user groups as shown in the Figure \ref{fig:static}

\subsubsection{Visual Exploration of Static Clustering}

To simultaneously depict the overall differences between different clusters and the distribution within each cluster, this study proposes a hybrid visualization approach that combines radar charts and violin plots for visualizing player clustering results. Radar charts provide a direct and visually appealing way to compare objects with multiple feature variables, while violin plots combine box plots and density plots to display data summary statistics and distribution shapes.

\paragraph{Visualization Design}
In the design of the visualization, the axes of the radar chart precisely correspond to the box plots in the violin chart, spanning the entire violin chart. Each violin plot represents the distribution of a feature value within a cluster. Each player's features are evenly mapped to a portion of the polar coordinates in the radar chart, and the violin plots with the same feature are arranged along the corresponding arc of the radar chart in the order of clusters. Additionally, different colors are used to fill the violin plots corresponding to different clusters, enabling the comparison of differences between different clusters for the same feature, differences between different features within the same cluster, and the distribution of data for the same feature within the same cluster. As the box plots in the violin chart are displayed in polar coordinates, the "box" also exhibits certain polar characteristics, such as the longer length of the side representing the upper quartile (i.e., the larger radius) compared to the side representing the lower quartile (i.e., the smaller radius).

\paragraph{Interaction Design}
The visualization incorporates user-friendly interactive design, allowing users to explore the data through intuitive interactions. When the user hovers their mouse over the box plot within the violin chart, statistical data pertaining to the user's features within that cluster are displayed, including the median, quartiles, maximum, and minimum values. Furthermore, users can select different clusters for comparison by clicking on the options in the top left corner. The more clusters selected, the smaller the portion of the radar chart occupied by each individual violin plot, resulting in a more coarse depiction of the data distribution shape. Conversely, selecting fewer clusters results in a larger portion of the radar chart occupied by each individual violin plot, providing a more detailed portrayal of the data distribution shape. Users have the flexibility to compare the differences in various features within the same cluster by selecting only one cluster, or to meticulously compare the clusters of interest in pairs.

\paragraph{Case Study}
Upon analyzing the results of the previous time series clustering, it is evident that many user features experienced significant changes in the third week. Thus, a deep clustering analysis was conducted on the third week using static user features, resulting in five distinct clusters. These clusters were visualized using the aforementioned visualization view, focusing on six features that reflect user social interaction, activity level, gaming proficiency, and willingness to make in-game purchases: segment, level, online time, average survival time, number of intimate friends, and weekly diamond increment. Due to the substantial differences in feature scales, all features of each player were normalized to the range [0, 1] during data processing, enabling a clear comparison of the player levels within the overall player population across different clusters.

The visualization results are shown in Figure 11. It can be observed that the majority of players have relatively low online time, which could be attributed to a small number of highly active players with longer online time, thus inflating the average level. Similarly, it can be noted that most players have a very low increment in diamonds, which aligns with direct observations of the data, indicating that the majority of players do not make any in-game purchases, while a few players make small purchases, and a very small number of players make high-value purchases. As for the other features, some differences are evident among the various clusters.

\begin{figure}[H]
    \centering
    \includegraphics[width=0.8\textwidth]{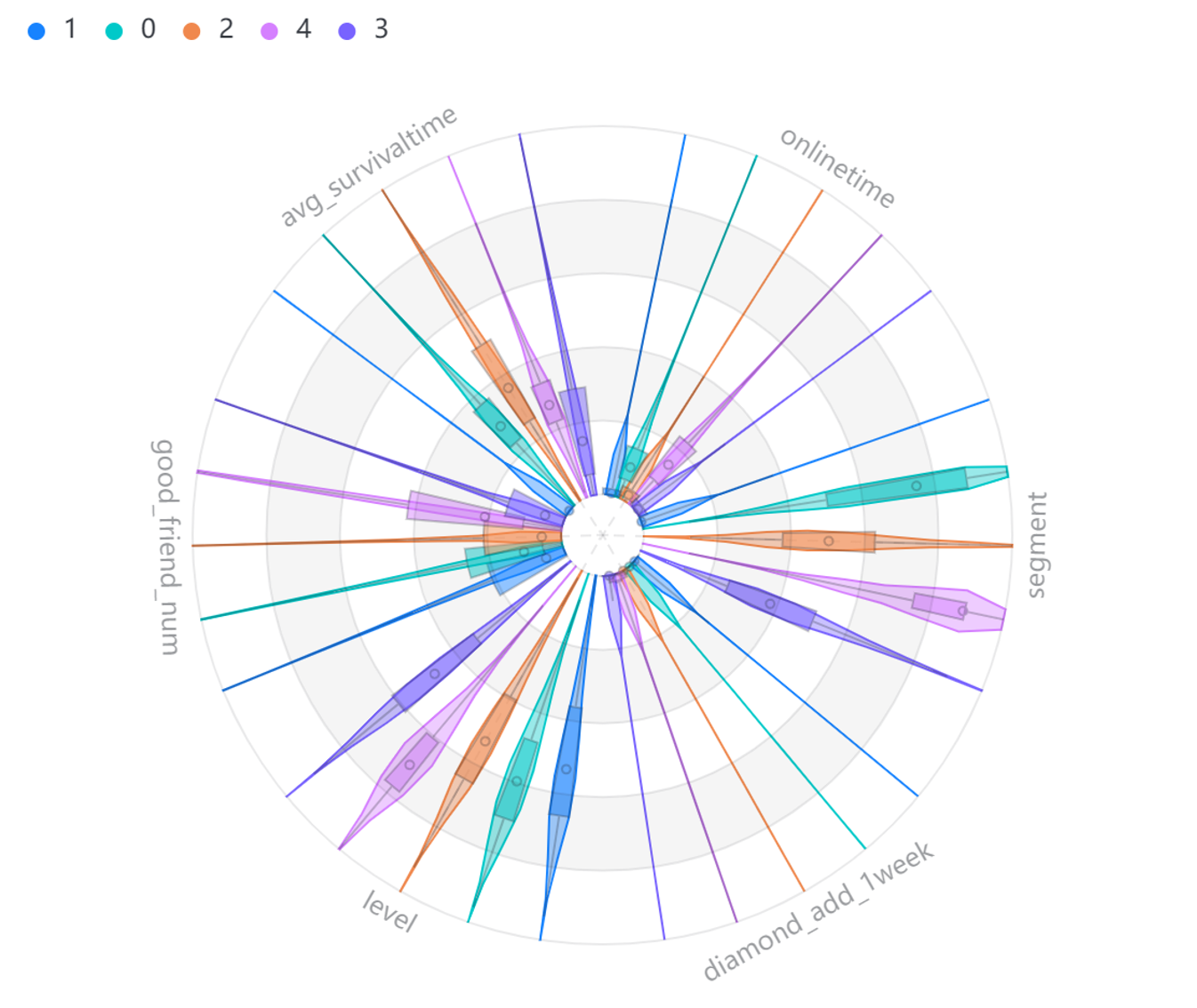}
    \caption{Visualization overview of static clustering.}
    \label{fig:Visualization overview of static clustering.}
\end{figure}

Comparing the different clusters, Cluster 4 exhibits the highest levels across all features except for average survival time. Users in this cluster have higher ranks and online time, indicating their loyalty to the game. Additionally, their higher segments and average survival time suggest that they possess strong gaming skills and their own understanding of the game, acquired through extensive gameplay. The higher number of intimate friends also indicates that due to their high game activity and strong gaming skills, they have accumulated a considerable number of friends and developed close relationships with them.

On the other hand, Cluster 2 has the longest average survival time, surpassing even the players in Cluster 4. However, the levels of other features are relatively lower. Analyzing the players in Cluster 2, it is possible that they have a gameplay strategy that leans more towards survival and "chicken" victories, indicating a preference for a lone wolf playstyle. In contrast, players in Cluster 4 are more enthusiastic about killing enemies and showcasing their skills, rather than solely focusing on their own survival. This difference reflects variations in gameplay strategies and preferences among players.

\begin{figure}[H]
    \centering
    \includegraphics[width=0.6\textwidth]{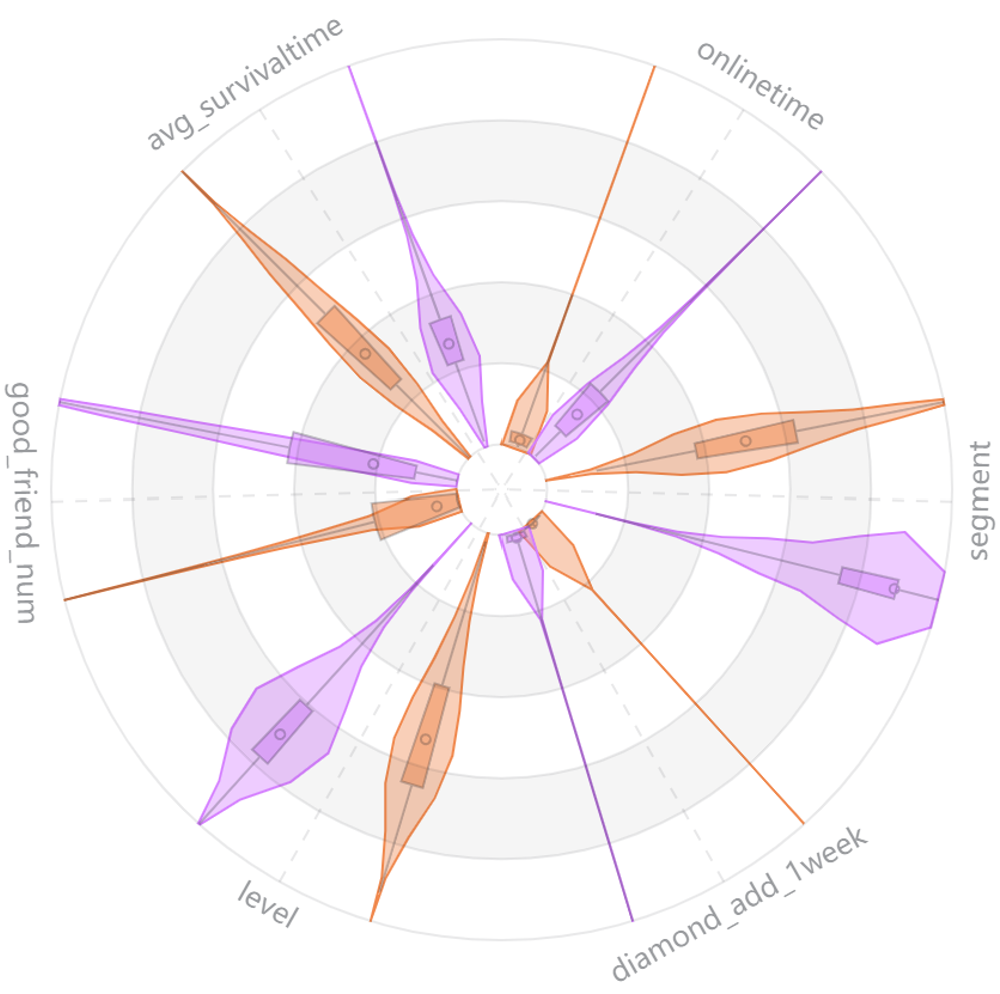}
    \caption{Comparison between cluster 4 and cluster 2.}
    \label{fig:Comparision between cluster 4 and cluster 2.}
\end{figure}

In stark contrast to Cluster 4, Cluster 1 exhibits the lowest levels across all user features, showing significant differences from the other clusters, indicating that they are inactive users. However, it is worth noting that players in Cluster 1 still have a certain number of close friends and levels, indicating that these players were once active within the game but had minimal gameplay during the analyzed week. This suggests that they might be new players who only briefly explored the game or former players who have discontinued their engagement.

Players in Cluster 3 are similar to those in Cluster 1. Although they have shorter online time, they have achieved certain segments and survival times within the week. However, their levels are not high. This indicates that they possess some level of gaming skills and could be new players who have recently started playing the game without becoming deeply involved, or they could be former players who did not have sufficient time to play during the analyzed week.

\begin{figure}[H]
    \centering
    \includegraphics[width=0.6\textwidth]{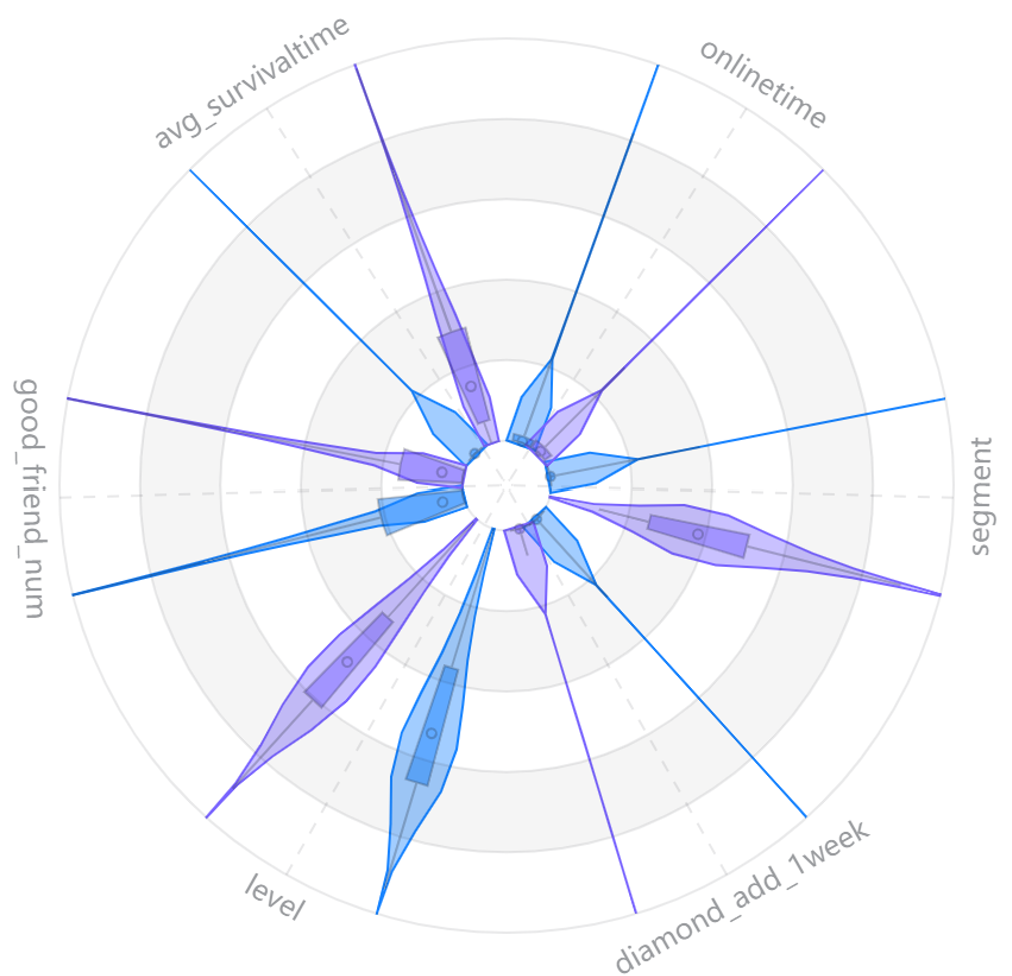}
    \caption{Comparison between cluster 1 and cluster 3.}
    \label{fig:Comparison between cluster 1 and cluster 3.}
\end{figure}

Cluster 0 exhibits similarities to Cluster 4 in various features, with some user characteristics ranking just below those of the active player cluster (Cluster 4), indicating a subgroup of moderately active players. By individually selecting these two clusters and carefully comparing their statistical information, it is revealed that the spending level of players in Cluster 4 is significantly higher than that of players in Cluster 0. All players in Cluster 4 are paying players, while only a small number of players in Cluster 0 make purchases. Additionally, a notable difference in the number of intimate friends between the two clusters is observed, leading to a reasonable inference that there exists a strong correlation between the willingness to make in-game purchases and the number of game friends, particularly intimate friends.

\begin{figure}[H]
    \centering
    \includegraphics[width=1\textwidth]{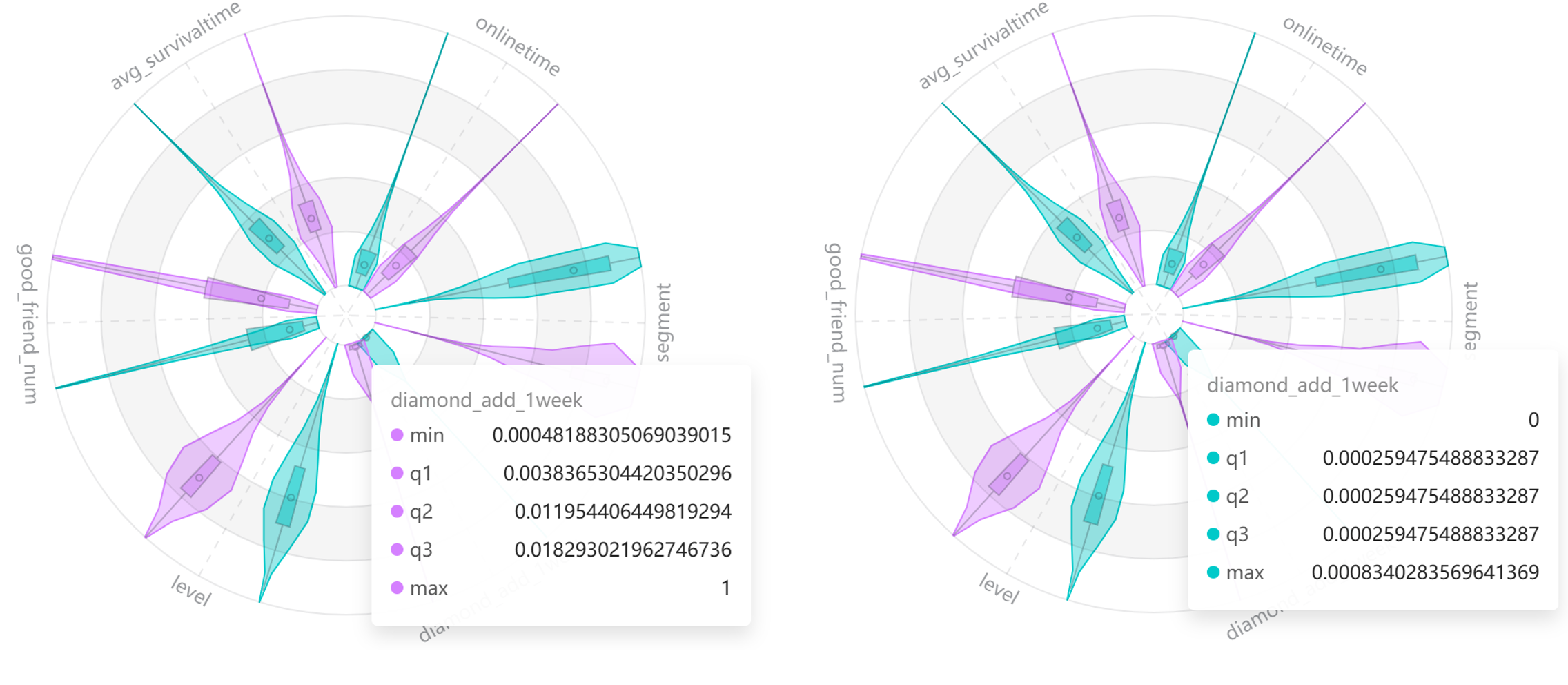}
    \caption{Comparison between cluster 0 and cluster 4.}
    \label{fig:Comparison between cluster 0 and cluster 4.}
\end{figure}

\subsubsection{Analysis of User Groups on Network Attribute Distribution}
Through the above clustering, we get 5 user groups, and the various internal interactions of these 5 user groups may still exist. Considering that network properties within different user groups can help us analyze the tightness of these subgraphs, we chose the following metrics:

\begin{itemize}

\item \textbf{Connected Components(CC): }Connected Components refer to the largest subsets of an undirected graph where every two vertices are connected, meaning there exists a path between any pair of vertices within the subset. The large number of connected components indicates that the graph is composed of several relatively independent subsets that are not directly connected. More independent connected components may mean that the system is more vulnerable.
\item \textbf{Average Clustering Coefficient(ACC): }The Average Clustering Coefficient, also known as the network's clustering coefficient, is a measure used in network analysis to quantify the degree to which nodes in a graph tend to cluster together. It provides insight into the cliquishness or local density of connections within a graph, reflecting how tightly knit the neighborhoods of individual nodes are. This coefficient ranges from 0 to 1, where a value close to 1 indicates a high tendency for nodes to form triangles (high local clustering), and a value close to 0 suggests that nodes do not tend to form tight clusters.

\item  \textbf{Triangles: }Number of Triangles refers to the number of triangles in an undirected graph, that is, the number of complete subgraphs formed by three nodes connected by three sides. Triangle is an important feature in network analysis, which can reflect the tightness and clustering characteristics of the network.

\end{itemize}

\begin{table}[h]
\centering
\caption{Network attribute analysis}
\label{tab:net}
\begin{tabular}{lccccc}
\hline
\textbf{Cluster} & \textbf{Node} & \textbf{Edge} & \textbf{CC} & \textbf{ACC} & \textbf{Triangles} \\
\hline
0 & 94241 & 205060 & 9505 & \textbf{0.1588} & 191370\\
1 & 46977 & 21744 & 27252 & 0.0371 & 5544\\
2 & 53288 & 27128 & 28483 & 0.0289 & 3867\\
3 & 4162 & 155 & 4001 & \textbf{0.0029} & 12\\
4 & 6726 & 2481 & 4498 & 0.0511 & 801\\
All & 205281 & 1207728 & 2849 & 0.1470 & 606735\\
\hline
\end{tabular}
\end{table}

\begin{figure}[H]
    \centering
    \includegraphics[width=0.6\textwidth]{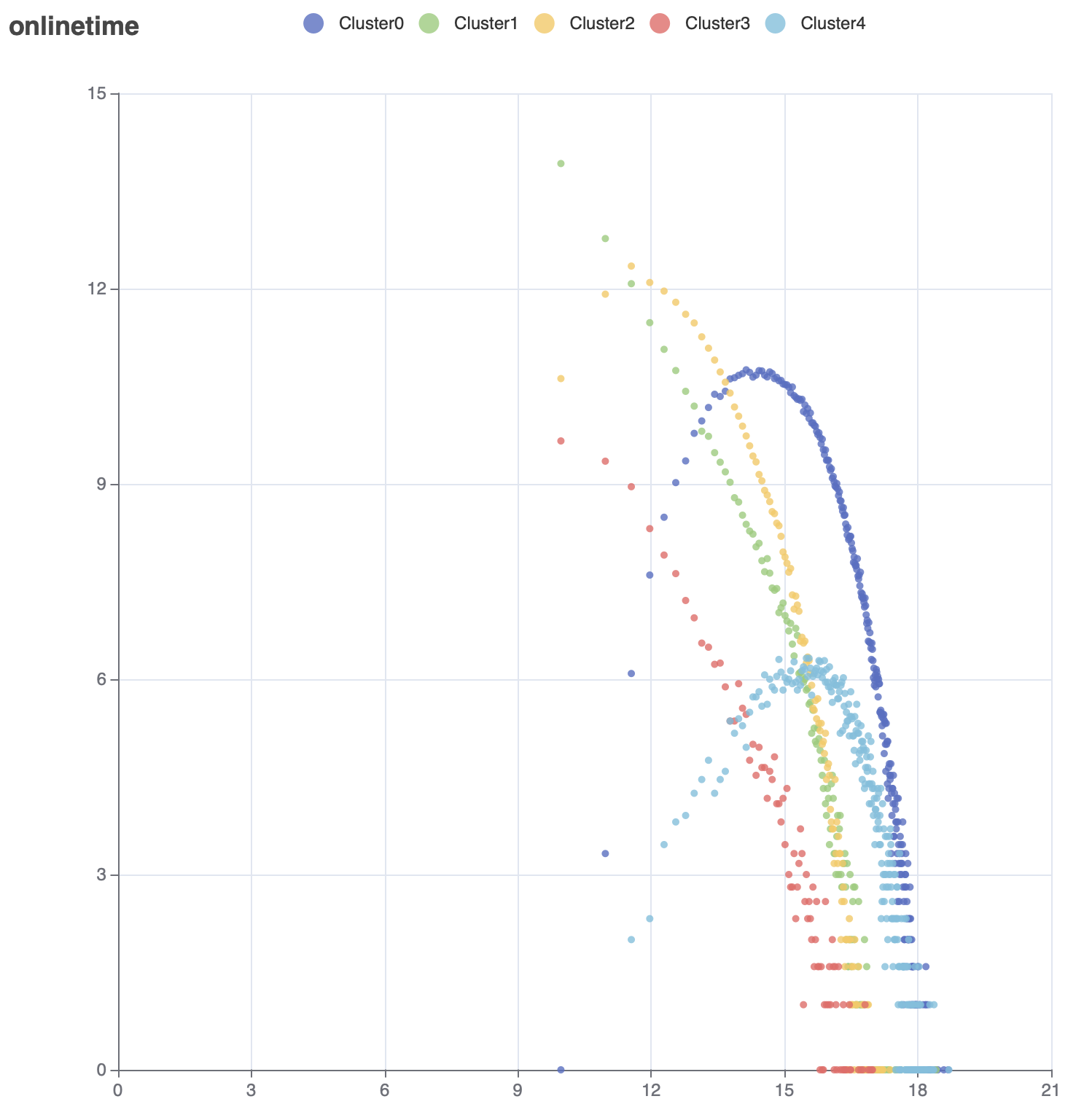}
    \caption{Online duration feature visualization.}
    \label{fig:net-on}
\end{figure}

As shown in the Table \ref{tab:net}, Clutser 1 has the highest ACC value among all user groups, which means that Clutser 1  has the tightest internal structure, while Clutser 3 is the most dispersed and vulnerable. Meanwhile, since the total graph structure containing all users has the most nodes and the least number of interconnection components, there are more connections between different user groups. Combined with the online duration distribution in \ref{fig:net-on}, it can also be found that the tighter the internal structure of the user group, the longer the online duration. Based on the above analysis, we can boldly speculate that if we want to further stimulate the activity of game players, the planning department can add more user interaction design in the future event planning.

\bibliographystyle{alpha}
\bibliography{sample}
\end{CJK}
\end{document}